\documentclass[a4paper,11pt]{article}
\usepackage[a4paper, total={17cm, 25cm}]{geometry}
\usepackage{fancyhdr}

\usepackage{graphicx}
\usepackage{epstopdf}
\usepackage{amsmath}
\usepackage{graphicx}
\usepackage[english]{babel}
\usepackage{subfig}
\usepackage{amssymb}
\usepackage{wrapfig}
\usepackage{pdflscape}
\usepackage{lscape}
\usepackage{longtable}
\usepackage{appendix}
\usepackage{textcomp}
\usepackage{multirow}
\usepackage{array}
\usepackage{txfonts}

\fancypagestyle{plain}{
\fancyhf{} 
\fancyhead[RH]{\thepage}
\fancyfoot[CF]{28th Texas Symposium on Relativistic Astrophysics \\ Geneva, Switzerland -- December 13-18, 2015}

}
\begin{document}

\title{\bf Narrow-line Seyfert 1 galaxies \\ -- rebels of the AGN family}

\author{
E. J\"{a}rvel\"{a}, Aalto University Mets\"{a}hovi Radio Observatory, 02540 Kylm\"{a}l\"{a}, Finland \\ and Aalto University Dept of Radio Science and Engineering, FI-00076 AALTO, Finland
\\
A. L\"{a}hteenm\"{a}ki, Aalto University Mets\"{a}hovi Radio Observatory, 02540 Kylm\"{a}l\"{a}, Finland \\ and Aalto University Dept of Radio Science and Engineering, FI-00076 AALTO, Finland
}

\date{\today}

\maketitle

\begin{abstract}

Narrow-line Seyfert 1 galaxies, with their extreme properties, defy our current knowledge of active galactic nuclei 
and relativistic jet systems. They excite, and might help us answer, many questions concerning the evolution and unification
of AGN, but still remain a poorly studied class of AGN as such. 

We did an extensive study of a large sample of NLS1s using various statistical methods, for example, multiwavelength 
correlations and principal component analysis. We wanted to examine how and where in NLS1s different kinds of radiation are produced,
and how the emission properties are connected to other intrinsic AGN properties. In addition we present the early results of our ongoing
research about the large-scale environments of NLS1s. We also introduce the Mets\"{a}hovi Radio Observatory NLS1 survey and its first results,
and show some early results for individual sources.

\end{abstract}


\section{Introduction}

The presence of powerful relativistic jets in narrow-line Seyfert 1 galaxies (NLS1) was confirmed when Fermi detected 
gamma-rays from a handful of them. In the current active galactic nuclei (AGN) paradigm powerful relativistic jets are 
produced only in massive elliptical galaxies with supermassive black holes, but NLS1s challenge this scenario 
since they have lower black hole masses, higher accretion rates, preferably compact radio morphology and they reside mostly in spiral galaxies.

Due to Fermi's discovery the AGN unification schemes have to be revised to fit in NLS1s, the third gamma-ray emitting AGN class. 
It also invokes questions about the AGN evolution; what triggers and maintains the AGN activity, and what are the evolutionary lines 
of the different populations? NLS1s complicate the whole AGN scenario, but also offer a new perspective at the jet phenomena at a different scale.

NLS1s are a poorly studied class of AGN. It seems that a surprisingly large fraction of them are radio-loud 
and thus possibly host jets; but also some of them seem to be totally radio-silent. This, as well as other observational evidence, 
implies that they do not form a homogeneous class. However, we are not certain what is the origin of the radio loudness,
but, for example, the properties of the host galaxy and the large-scale environment might play a role.

To get more insight into the emission and other intrinsic properties of NLS1s we performed a multivariable study of a large sample
of NLS1s. This study is reviewed in Section~\ref{sec:mvstudy}. In Section~\ref{sec:largescale} we discuss the ongoing study of the 
large-scale environments of NLS1s, and in Section~\ref{sec:radioobs} we present the Mets\"{a}hovi NLS1 survey. Finally, in
Section~\ref{sec:discussion} we summarise the results and the ongoing research.

\section{Multivariable study of NLS1s}
\label{sec:mvstudy}

\subsection{Sample and data}

We did an extensive statistical study of a large sample of NLS1s \cite{2015jarvela1} in which our aim was to estimate via which processes and where the various 
kinds of emission are produced in them. We did this by compiling multiwavelength and spectral data from literature for a sample of NLS1s.
With this data set we explored the connections between different wavebands, and spectral and other intrinsic AGN properties using various correlation 
analyses, for example, Pearson's product-moment and Spearman's rank correlations, and principal component analysis.

Using three references \cite{2006zhou1, 2008yuan1, 2006komossa1} we selected sources which had radio data from 
the Very Large Array (VLA) Faint Images of the Radio Sky at Twenty-Centimeters (FIRST) survey. Our final sample consists of 292 NLS1s.
All the data were gathered from publicly available archival sources, and thus are not simultaneous. Since NLS1s are generally variable multiwavelength
studies should ideally be performed with simultaneous data, but observations of NLS1s are too scarce to do this.

Radio data from FIRST, optical data from Sloan Digital Sky Survey (SDSS), and X-ray data from ROSAT All Sky Survey (RASS) were retrieved from ASDC{\footnote{http://www.asdc.asi.it/}}.
We have radio and optical data for all sources, and X-ray data for 109 sources. Infrared (IR) data from Wide-field Infrared Survey Explorer 
(WISE) All-Sky Source Catalog \cite{2012cutri1} were retrieved from the NASA/IPAC Infrared Science Archive (IRSA{\footnote{http://www.irsa.ipac.caltech.edu}}) and we have it
for 282 sources.

\subsection{Radio loudness} 

We computed radio loudness ($RL$) values using the commonly defined $RL$ value; the ratio of 1.4 GHz radio flux density ($F_{\text{R}}$) and 440 nm optical flux density ($F_{\text{O}}$).
We used the K-corrected \cite{2011foschini1} radio flux densities from the FIRST survey and K-corrected B-band optical flux density 
calculated using SDSS $u$- and $g$-band magnitudes. The radio loudness value distribution of our sample is shown in Figure~\ref{fig:rl-dist}. We divided our sample to 
subsamples by radio loudness; 195 sources with $RL >$ 10 form the radio-loud (RL) subsample, and 97 sources with $RL <$ 10 the radio-quiet (RQ) subsample.

\subsection{Black hole mass} 

Black hole masses were estimated using the FWHM(H$\beta$) -- $L_{\text{5100}}$ (the monochromatic luminosity at 5100\AA{}) mass scaling relation \cite{2005greene1}.
This method does not take into account possible orientation effects caused by the geometry of the broad-line region of the source and the viewing 
angle \cite{2011decarli2}, but methods based on $\sigma_{\ast}$ or $L_{\text{bulge}}$ can not be used due to the lack of data.

\begin{figure*}[ht!]
\centering
\begin{minipage}{0.45\textwidth}
\centering
\includegraphics[width=1\textwidth]{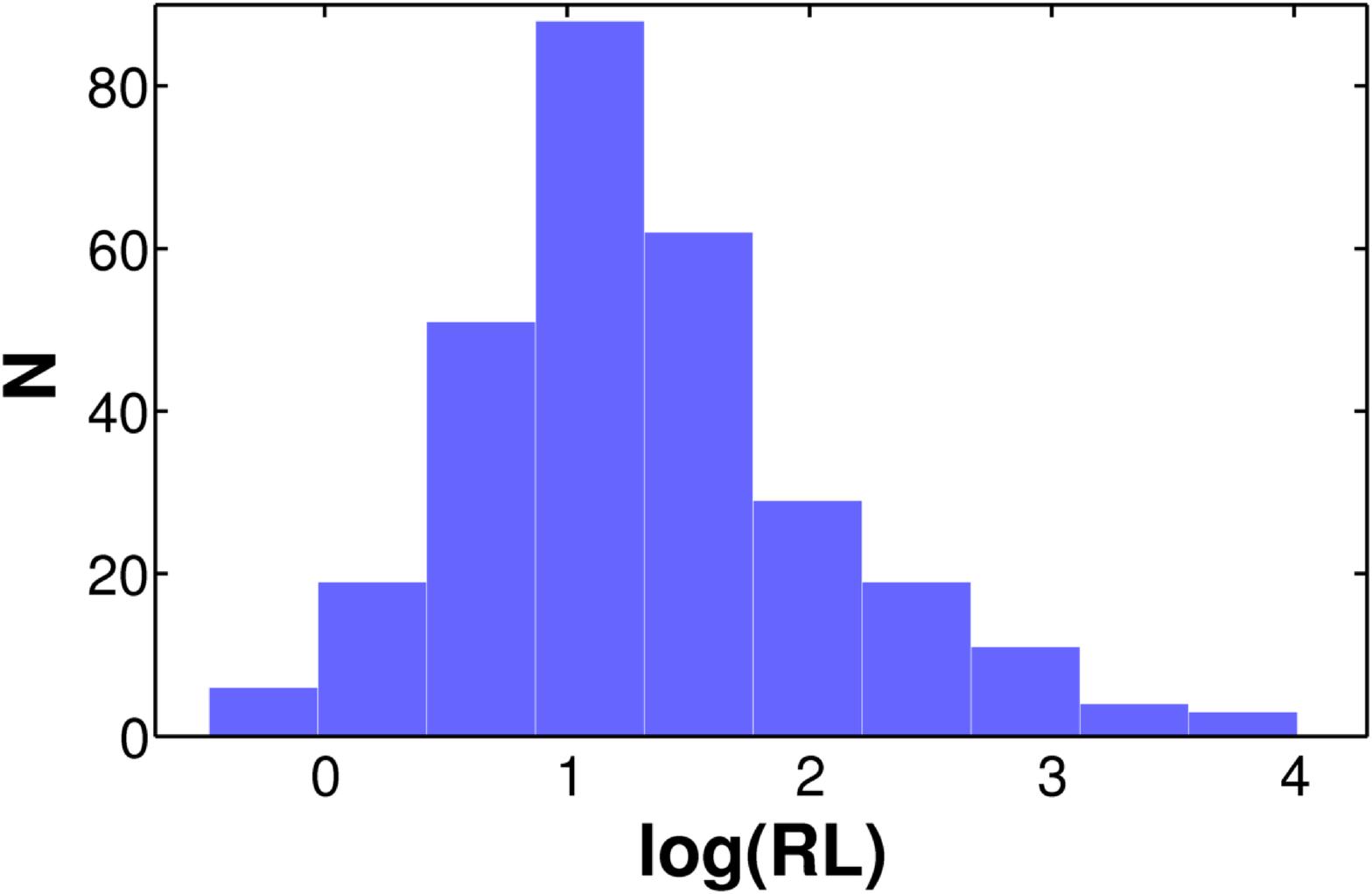}
\caption{Logarithmic radio loudness distribution of our whole sample.} \label{fig:rl-dist}
\end{minipage}\hfill
\begin{minipage}{0.49\textwidth}
\centering
\includegraphics[width=1\textwidth]{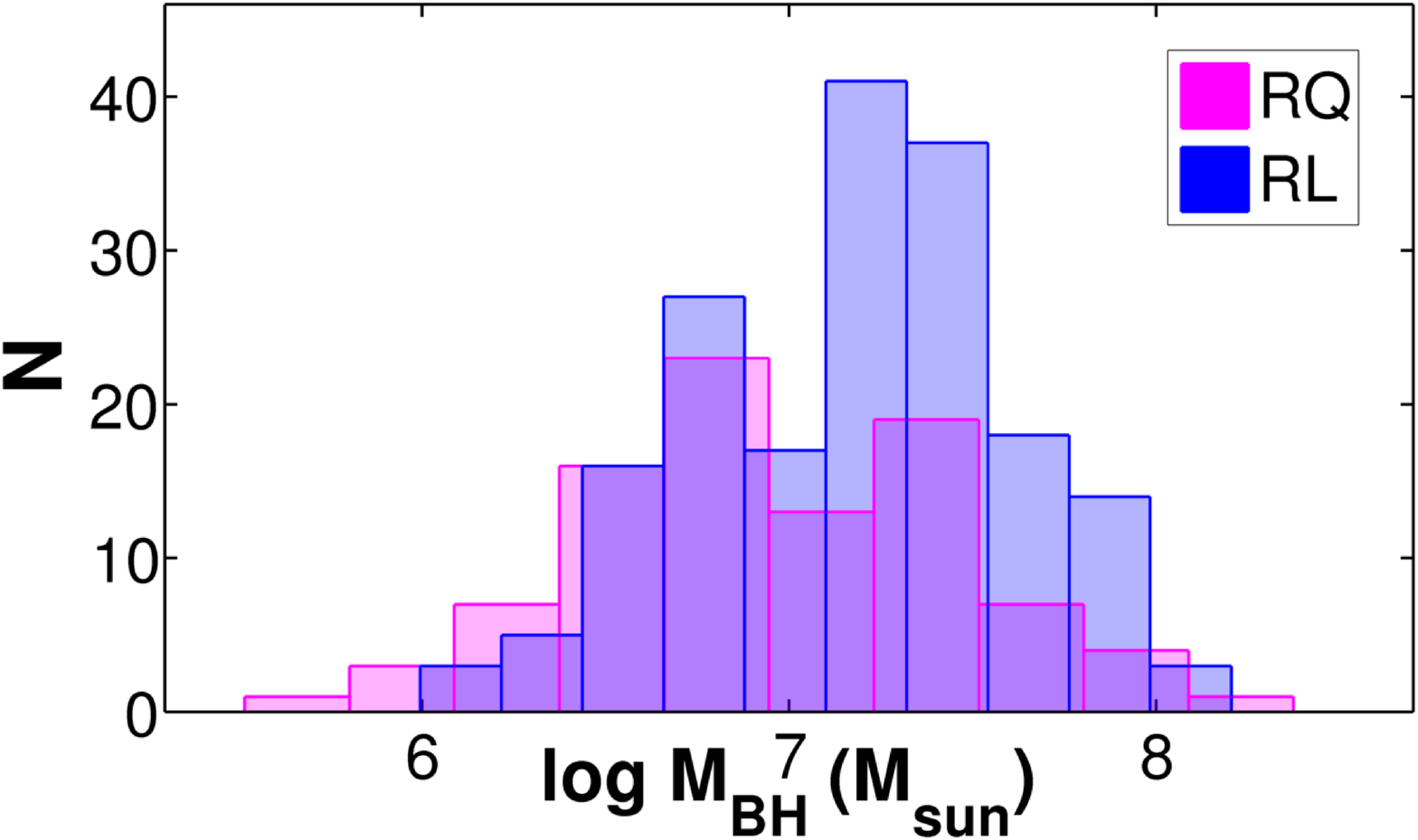}
\caption{Logarithmic black hole mass distributions of RQ and RL subsamples.} \label{fig:mbh-dist}
\end{minipage}
\end{figure*}

We were able to estimate the $M_{\text{BH}}$ for 275 sources. The $M_{\text{BH}}$ distributions for RQ and RL subsamples are shown in Figure~\ref{fig:mbh-dist}.
The Kolmogorov-Smirnov (K-S) test for $M_{\text{BH}}$ suggests that the RQ and RL subsamples are not drawn from the same distribution (p=2.04$\times10^{-4}$). 
On average RL sources have more massive black holes than RQ sources.

\subsection{Multiwavelength correlations}

We calculated multiwavelength correlations using both, flux densities and luminosities. The used methods were Pearson's product-moment and Spearman's rank correlations -- normal, 
and partial, with $z$ as the third variable. We calculated the correlations separately for RQ and RL subsamples. Results from all the correlations were consistent with each other. 

The correlations suggest that in RL sources the predominant sources of radio, optical, and X-ray emissions are the relativistic jets. 
The origin of infrared emission remains unclear, but it is probably a mixture of both, thermal and non-thermal, processes. RQ sources do not host a jet, or the jet is very weak. 
Radio and IR emission in them is more likely generated via star formation processes, and the optical and X-ray emission originate in the inner parts of the AGN. 

\subsection{Principal component analysis}

Principal component analysis is a statistical method which can be used to determine which properties cause most of the variability in a data set. 
We performed PCA with multiwavelength data, $M_{\text{BH}}$, R4570, and FWHM(H$\beta$). \emph{Eigenvector 1} accounts for 34\% -- 37\% of the variance and distinguished 
between the $M_{\text{BH}}$, and the brightness at IR and optical. X-ray emission is connected to IR and optical but contributes less. 
Radio emission is rather insignificant except in the RQ subsample where it is connected with IR and optical, confirming the suggestion of its stellar 
origin in these sources. The prominence of $M_{\text{BH}}$ implies that EV1 might be similar to EV2 in \cite{2004grupe1}. \emph{Eigenvector 2} is 
responsible for 19\%-24\% of the variance and clearly distinguishes between R4570 and FWHM(H$\beta$) in all samples. IR and optical seem to be 
connected to FWHM(H$\beta$) to some extent. EV2, dominated by the anticorrelation of R4570 and FWHM(H$\beta$), is similar to the traditional EV1 \cite{1992boroson1, 2002boroson1, 2012xu1}.

Eigenvectors in our study seem to be reversed compared to previous studies, but as pointed out in \cite{2004grupe1}, each sample has its own eigenvectors that 
depend on the variables used and their range. Our results can be explained taking into account that AGN form a continuum in which the properties change from one
class to another, as well as inside a class, and that our study is the first one using a pure NLS1 sample.
When using mixed samples the properties that change from one source type to another create the EV1, which is transformed to a sequence of properties within a 
pure sample but in a less significant role (EV2 of our NLS1 sample). EV1 of our NLS1 sample describes the properties that cause most of the differences within 
that source type only. Previous studies have mostly used only optical data, thus for consistency, we tested our analysis by performing it with optical data only.
This, however, did not change our results.

\section{Large-scale environment}
\label{sec:largescale}

Host galaxy and close-by and large-scale environment studies of NLS1s will improve our knowledge of the properties of them as a class, as well as their
evolution and the AGN unification in general. Interesting questions are, for example, whether NLS1s are different to broad-line Seyfert 1 galaxies (BLS1) in host 
or environmental properties, and are Type 1 and Type 2 AGN similar or not.

The differences between the host galaxies of NLS1s and other types of Seyfert galaxies (BLS1s, Sy2s) are prominent. Both, NLS1s and BLS1s are preferably 
hosted by disk-like galaxies, but in NLS1s the incidence of bars is higher, they exhibit more star formation \cite{2003crenshaw1, 2006deo1, 2007ohta1, 2010sani1} 
and are less likely to be merging/interacting \cite{2007ohta1}. The absence of interaction suggests that the growth and evolution of NLS1s has been 
and is dominated by secular processes \cite{2011orbandexivry1} and the evolutionary path differs from that of other Seyferts.

Environmental studies point to Sy1s and Sy2s residing in different environments; Sy2s have more close companions, but reside in less dense large-scale 
environments and are generally of later type, whereas Sy1s have less close companions, reside in denser large-scale environments and are more 
evolved \cite{1999dultzinhacyan1, 2001krongold1, 2006koulouridis1}. \cite{2001krongold1} and \cite{2014ermash1} did not find differences between the environments of 
NLS1s and Sy1s, but \cite{2001krongold1} found that compared to non-active galaxies with similar properties NLS1s have fewer companions and are farther away 
from them than non-active galaxies. They also found that NLS1 host galaxies are smaller than hosts of Seyfert 1 or normal galaxies. 

\cite{2011lietzen1} used SDSS DR7 to construct a luminosity density field of luminous red galaxies to study the large-scale environments of different AGN classes.
They conclude that Sy1s and Sy2s reside in different environments. We used their density field data to study the large-scale environments of NLS1s;
we have density field data for 960 sources with $z <$ 0.4. We are interested in the environmental distribution of NLS1s, and whether it is different compared 
to other types of AGN. Table~\ref{tab:denspros} shows average densities for NLS1 subsamples and the fraction of sources in voids, filaments and superclusters. 
Other types of sources from \cite{2011lietzen1} are listed for comparison. We use the same definition for voids, filaments and clusters as in 
\cite{2011lietzen1}; in void density $<$ 1.0, in filament 1.0 $<$ density $<$ 3.0 and in supercluster density $>$ 3.0.

Based on the preliminary results the average density seems to increase with increasing $RL$ and the environmental distribution of sources seems to change with $RL$. 
However, K-S test suggests that the density distributions of our subsamples are similar. The average density for our whole sample is smaller than 
for any sample in \cite{2011lietzen1}, and the distribution to density bins is clearly different. There seems to be more NLS1s than Sy1s or Sy2s in voids, 
less in filaments and roughly the same fraction in superclusters. These results, along with a more extensive multivariable study with a bigger sample of 
NLS1s will be published in the near future.

\begin{table*}[ht]
\caption[]{Average density, percentage in voids, filaments and superclusters.}
\centering
\begin{tabular}{l l l l l l}
\hline\hline
                               & N     & Average density & Void (\%) & Filament (\%) & Supercluster (\%) \\ \hline
FR 1 galaxies                  & 793   & 3.01 $\pm$ 0.07 & 10        & 48            & 42                \\
Flat-spectrum radio galaxies   & 624   & 2.60 $\pm$ 0.07 & 13        & 55            & 32                \\
Seyfert 1 galaxies             & 1095  & 1.73 $\pm$ 0.04 & 34        & 51            & 15                \\
Seyfert 2 galaxies             & 2494  & 1.65 $\pm$ 0.03 & 35        & 52            & 13                \\
all NLS1s                      & 960   & 1.50 $\pm$ 0.04 & 44        & 44            & 12                \\
radio-silent                   & 799   & 1.48 $\pm$ 0.05 & 45        & 43            & 12                \\
radio-quiet                    & 63    & 1.52 $\pm$ 0.14 & 38        & 52            & 10                \\
radio-loud                     & 98    & 1.65 $\pm$ 0.13 & 42        & 43            & 15                \\ \hline

\end{tabular}
\label{tab:denspros}
\end{table*}

\section{Radio observations of NLS1s}
\label{sec:radioobs}

Radio observations are crucial for understanding the relativistic jet phenomena since the properties of the relativistic jets are somewhat directly 
connected to the radio emission properties in radio-loud AGN -- also in NLS1s. So far, NLS1s have been very scarcely observed at any, 
especially higher, radio frequencies. A few monitoring programs are targeting NLS1s (e.g. F-GAMMA{\footnote{http://www3.mpifr-bonn.mpg.de/div/vlbi/fgamma/fgamma.html}}), 
but they mainly concentrate on the brightest individuals. To properly study NLS1s as a class we need observations of bigger and more diverse samples.

\subsection{Mets\"{a}hovi pilot survey}

To amend the lack of radio observations we have launched an extensive observing program of NLS1s at 22 and 37 GHz at the Mets\"{a}hovi Radio Observatory,
Kirkkonummi, Finland, where a 14m radio telescope is used for monitoring large samples of AGN frequently. At the moment we have four NLS1 samples totaling up to 
145 sources chosen based on gamma-ray detections, radio-loudness, large-scale environment, and brightness on optical, IR or X-rays.
Monitoring of the NLS1 sample 1 started in February, 2012, of the sample 2 in November, 2013, and of the samples 3 and 4 in March, 2014, and are still ongoing.
Four NLS1s (1H 0323+342, SDSS 084957.97+510829.0, SDSS J094857.31+002225.4 and SDSS J150506.47+032630.8) were monitored already earlier, thus we have 
observations spanning much longer times for them. We aim to have at least three observations of each source separated by several months to see if they are 
at all detectable at 37 GHz and to examine the possible variability. For Mets\"{a}hovi observations we use S/N $>$ 4 as a detection limit. By the end of 
April, 2015, we had fulfilled this criterion for every source in samples 1 and 2. We were able to detect 15 of 78 sources (19.2\%), of which, unfortunately, 
seven sources only once. This might indicate that their flux densities are at our detection limit and/or that they are variable. 

\subsection{Additional radio data}

In addition to Mets\"{a}hovi data we have radio data from RATAN-600 and Owens Valley Radio Observatory.

RATAN-600{\footnote{http://www.sao.ru/ratan/}} is a radio telescope with a diameter of 576 m operated by the Russian Academy of Sciences. It is situated in 
The Special Astrophysical Observatory in Bolshoi Zelenchuk Valley of the Greater Caucasus in Russia. We have data from RATAN-600 for 29 NLS1s at frequencies 
4.8, 7.7, 8.2, 11.2 and 21.7 GHz. The observations were made during October, 2013, and March-April, 2014.

40m{\footnote{http://www.astro.caltech.edu/ovroblazars/index.php?page=home}} radio telescope based in the Owens Valley Radio Observatory (OVRO), California, USA, is used for 
monitoring a sample of AGN weekly. We have OVRO data at 15 GHz for 15 NLS1s between January, 2008 and the end of April, 2015. For OVRO observations we used S/N $>$ 4 as a detection limit.

Also Planck{\footnote{http://sci.esa.int/planck/}} data, between frequencies 30 and 857 GHz, for some NLS1s will be available in the future.

\subsection{Early results}

\begin{figure*}[ht!]
\centering
\begin{minipage}{0.33\textwidth}
\centering
\includegraphics[width=1\textwidth]{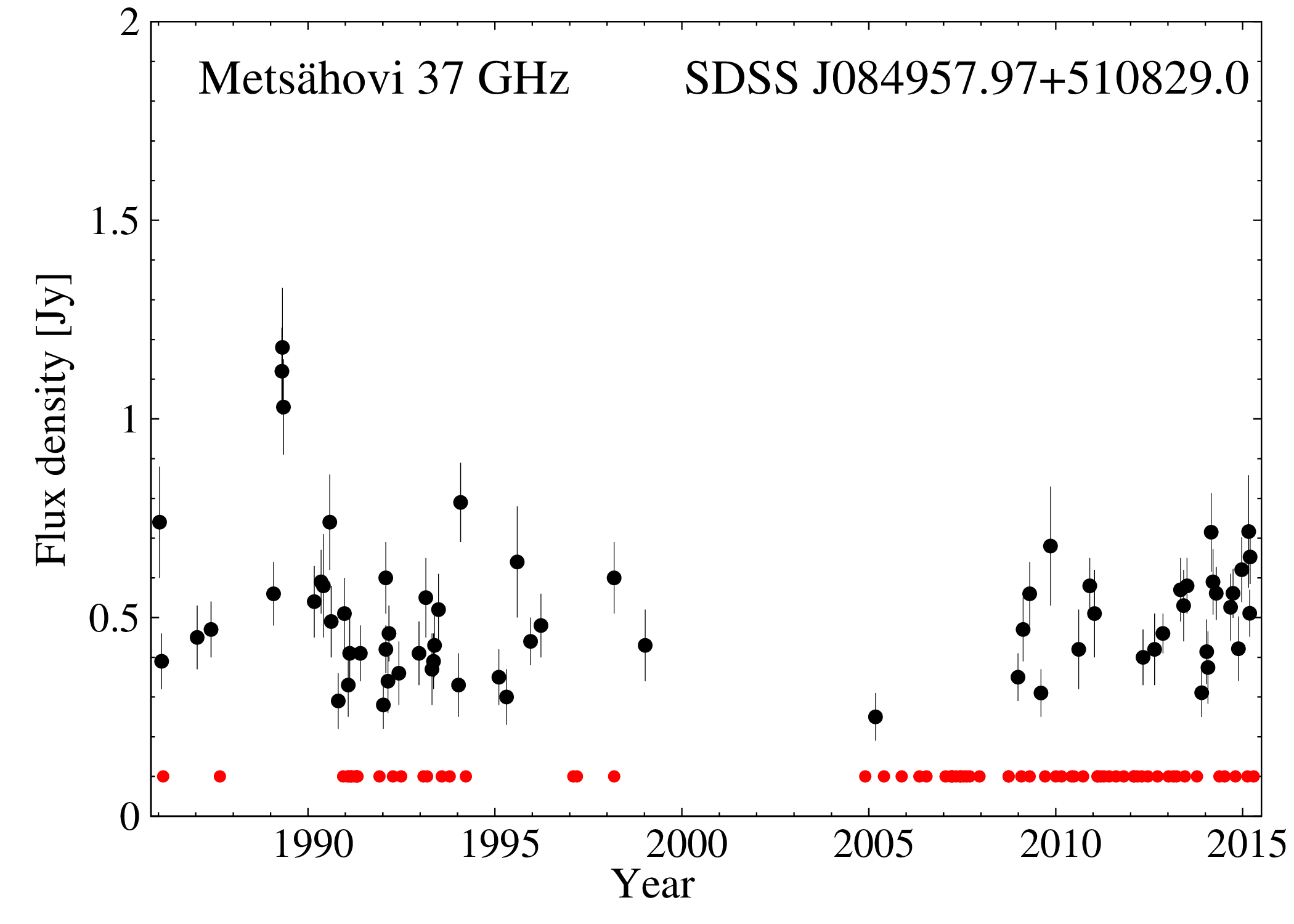}
\end{minipage}\hfill
\begin{minipage}{0.33\textwidth}
\centering
\includegraphics[width=1\textwidth]{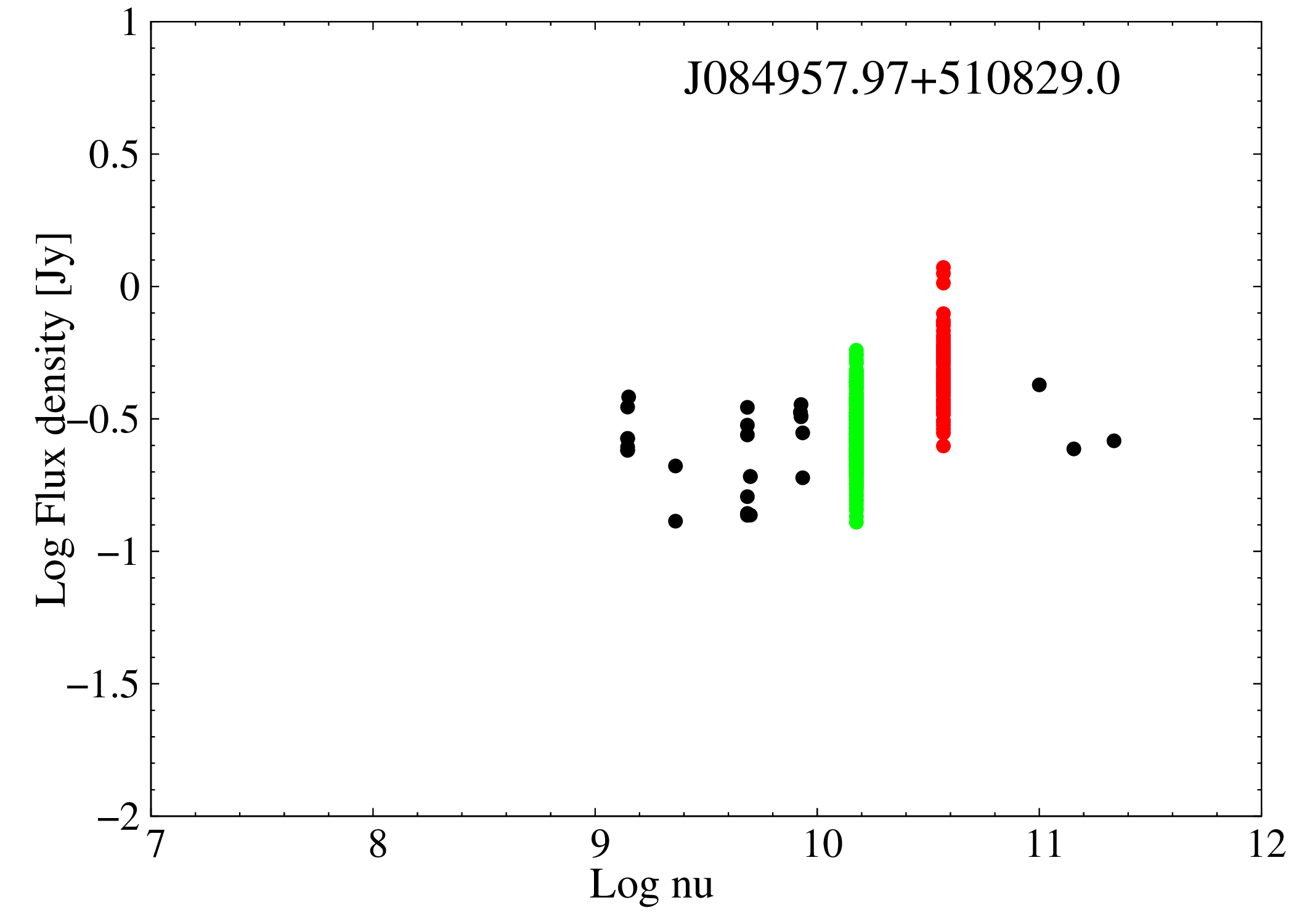}
\end{minipage}
\begin{minipage}{0.33\textwidth}
\centering
\includegraphics[width=1\textwidth]{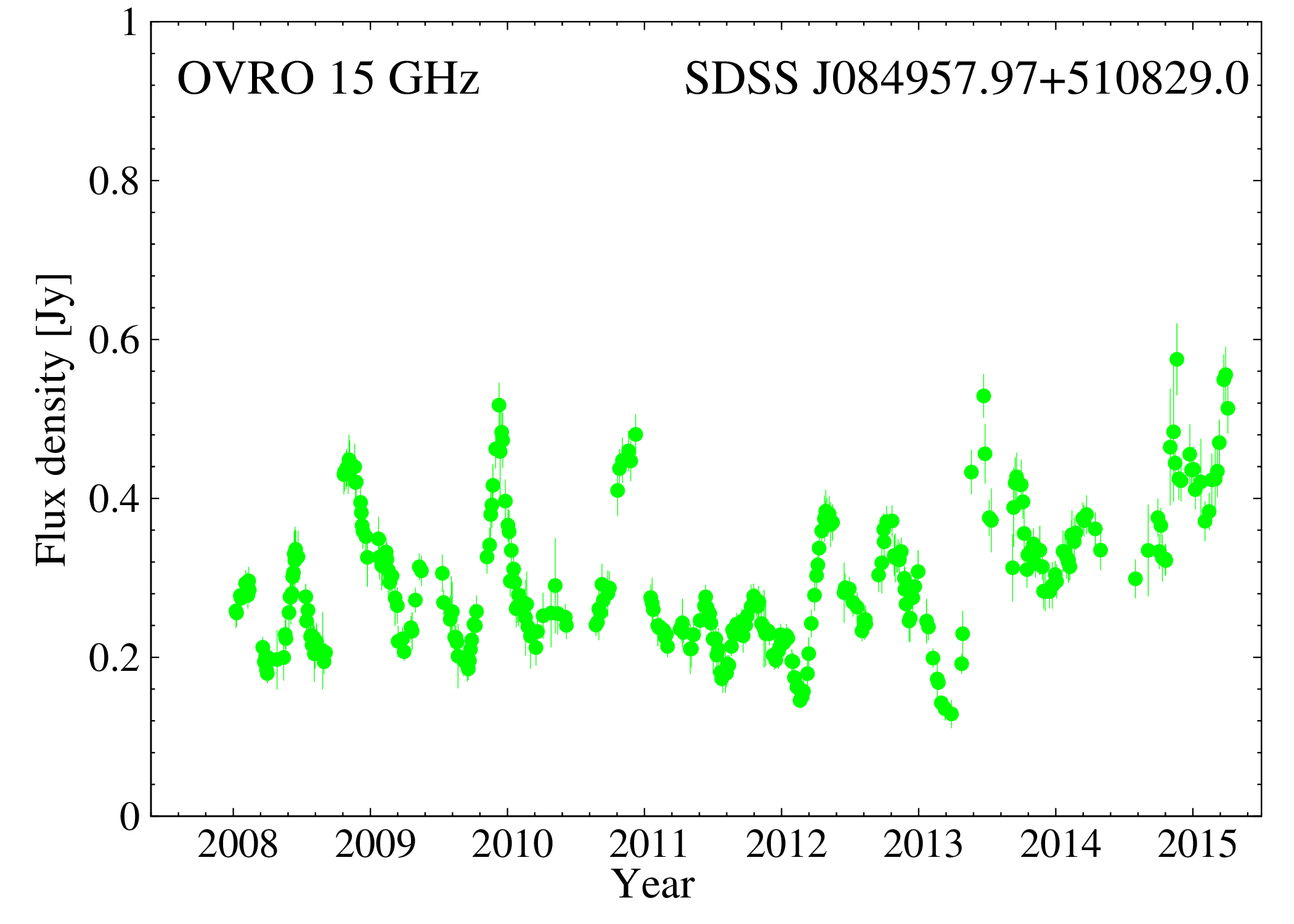}
\end{minipage}\hfill
\caption{SDSS J084957.97+510829.0. \emph{Left.} Mets\"{a}hovi light curve. Black circles are detections, and red circles are non-detections. 
\emph{Middle.} Radio spectrum. Black circles are archival data, OVRO -- green, Mets\"{a}hovi -- red, Planck -- blue.
\emph{Right.} OVRO light curve. All are detections. } \label{fig:0849}
\end{figure*}

Data for \emph{SDSS J084957.97+510829.0} is shown in Figure~\ref{fig:0849}. Mets\"{a}hovi light curve at 37 GHz is shown in the left panel (black circles are detections
and red circles are non-detections). This is the longest light curve we have for any NLS1 and it can be clearly seen that this source is variable. Radio spectrum with the archival 
(black) and new data (OVRO -- green, Mets\"{a}hovi -- red) is in the middle panel and implies that this source has a flat, somewhat variable, radio spectrum. The source is distinctly
variable at 15 GHz OVRO light curve (right panel, showing only detections).

\begin{figure*}[ht!]
\centering
\begin{minipage}{0.33\textwidth}
\centering
\includegraphics[width=1\textwidth]{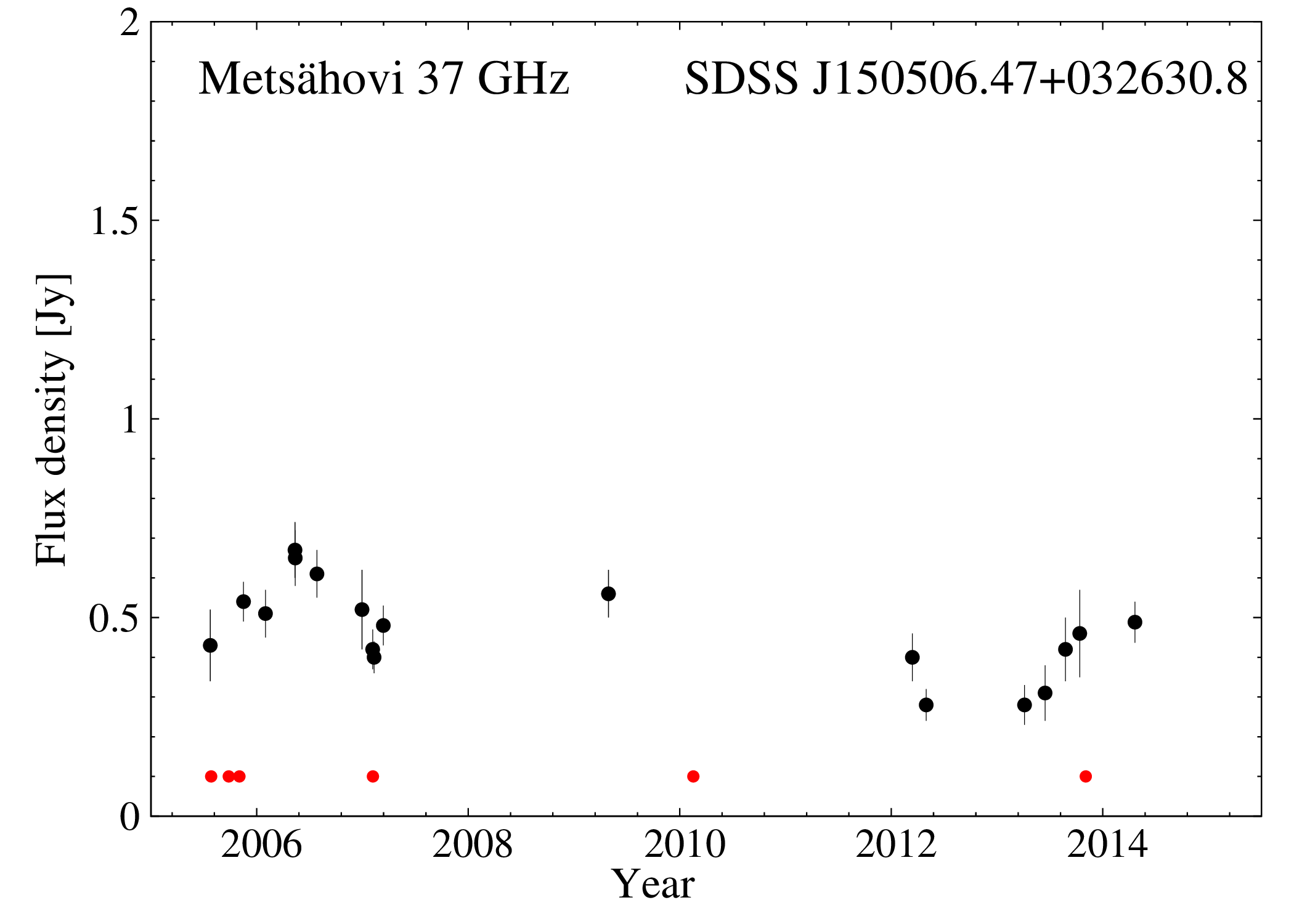}
\end{minipage}
\begin{minipage}{0.33\textwidth}
\centering
\includegraphics[width=1\textwidth]{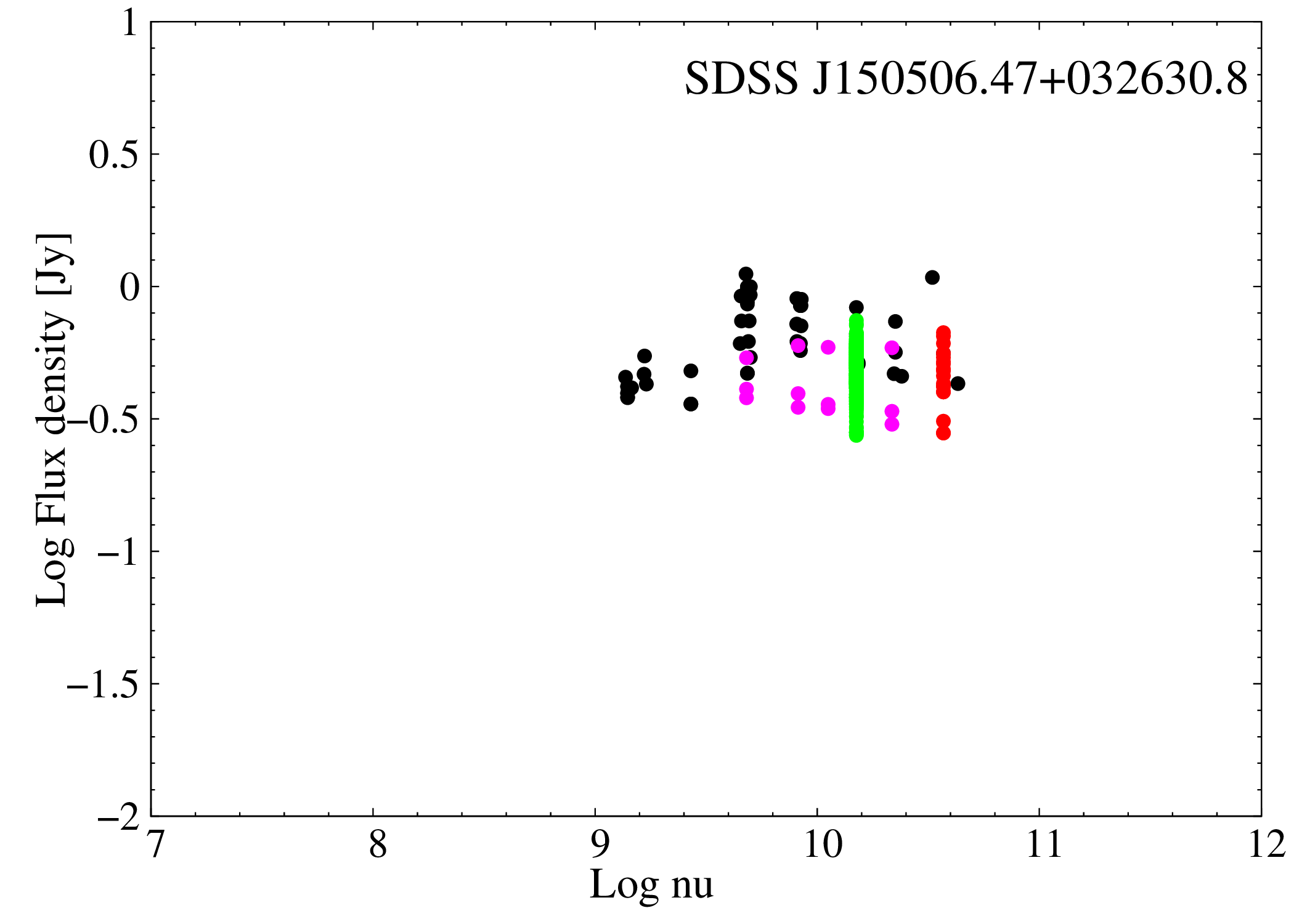}
\end{minipage}\hfill
\begin{minipage}{0.33\textwidth}
\centering
\includegraphics[width=1\textwidth]{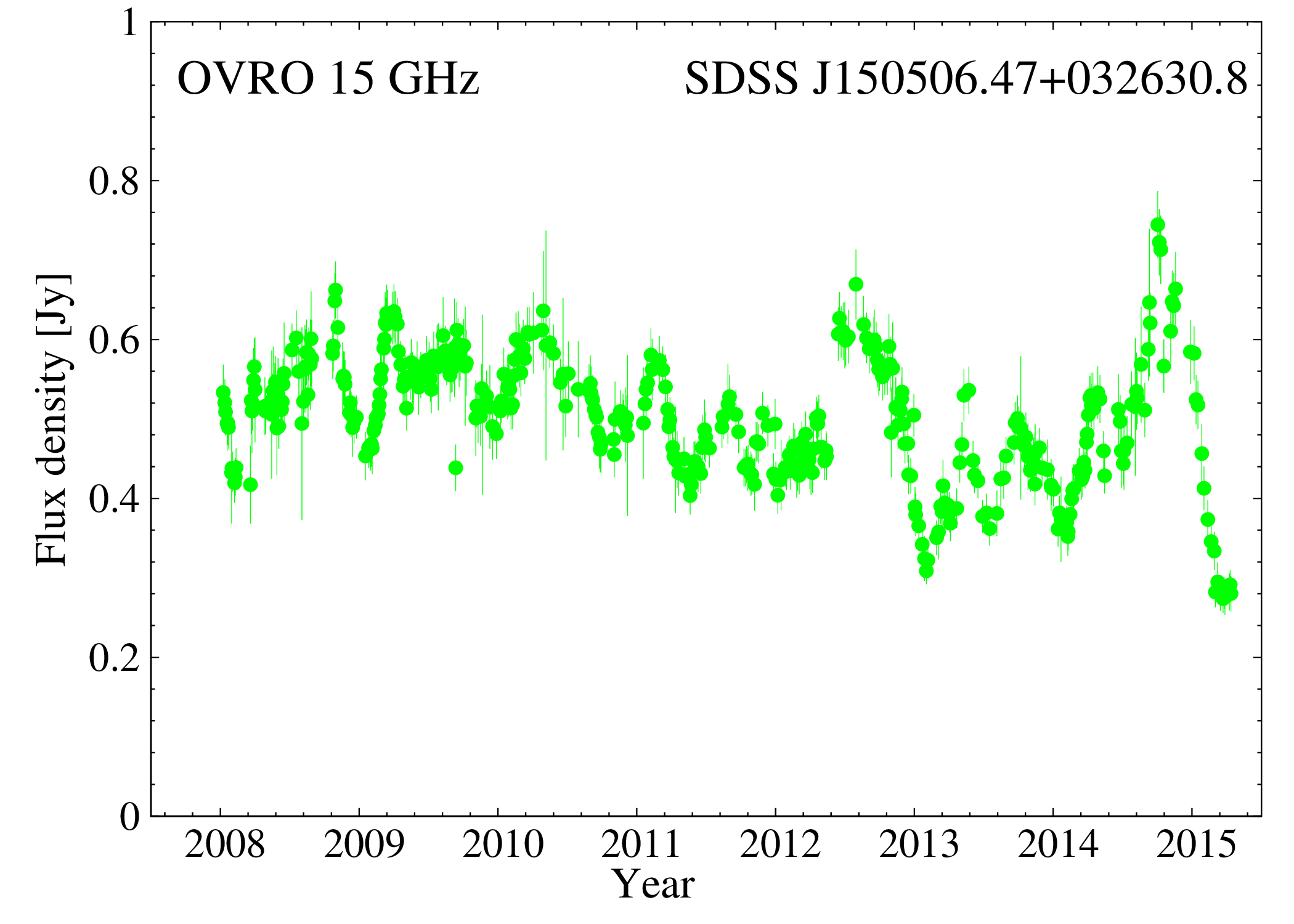}
\end{minipage}
\caption{SDSS J150506.47+032630.8. \emph{Left.} Mets\"{a}hovi light curve. Detections in black and non-detections in red. 
\emph{Middle.} Radio spectrum. Archival data -- black, RATAN -- magenta, OVRO -- green, Mets\"{a}hovi -- red.
\emph{Right.} OVRO light curve. All are detections.}  \label{fig:1502}
\end{figure*}

Figure~\ref{fig:1502} shows the data for \emph{SDSS J150506.47+032630.8}. Left panel shows the 37 GHz Mets\"{a}hovi light curve (detections in black and non-detections in red) which
indicates that the source is variable. The middle panel shows the radio spectrum (archival data -- black, RATAN-600 -- magenta, OVRO -- green, Mets\"{a}hovi -- red), which seems to
be flat and, to some extent, variable. The OVRO light curve at 15 GHz (only detections) in the right panel exhibits rather high amplitude variability.

Data and results for the whole pilot survey -- samples 1 and 2 -- will be published later this year. 

\section{Conclusions}
\label{sec:discussion}

Multivariable studies shed some light on the different emission mechanisms acting in RQ and RL NLS1s, but left some questions, for example, the origin of the IR emission,
unanswered. More detailed studies on this matter are needed to distinguish the contribution of star formation and AGN related emission processes in NLS1s, and to establish
the results at other wavebands.

Information about the large-scale environment of NLS1s and related AGN classes offers us a new tool to study the evolution and unification of AGN, and combined
with additional data helps us clarify the properties of NLS1s as a class. For example, the knowledge whether NLS1s are a homogeneous class effects how we 
interpret them and how they should be treated in the grand AGN scenario to effectively use them in studying the jet and AGN phenomena.

To properly address the issues concerning NLS1s additional observations of large, diverse samples at various wavebands are needed. Mets\"{a}hovi NLS1 survey,
being the first survey dedicated to systematical observations of NLS1s at high radio frequencies, will provide us with invaluable new information about this
interesting and unique class of AGN.

\end{document}